# The *ECLAIRs* micro-satellite mission for gamma-ray burst multi-wavelength observations

S. Schanne*, J.-L. Atteia, D. Barret, S. Basa, M. Boer, F. Casse, B. Cordier,
F. Daigne, A. Klotz, O. Limousin, R. Manchanda. P. Mandrou, S. Mereghetti,
R. Mochkovitch, S. Paltani, J. Paul, P. Petitjean, R. Pons, G. Ricker, G. Skinner

* CEA Saclay, DSM/DAPNIA/Service d'Astrophysique, F-91191 Gif-sur-Yvette, France,
tel: +33 1 69 08 15 47, e-mail: schanne @ hep.saclay.cea.fr

***Abstract:*** *Gamma-ray bursts (GRB) − at least those with a duration longer than a few seconds − are the most energetic events in the Universe and occur at cosmological distances. The ECLAIRs micro-satellite, to be launched in 2009, will provide multi-wavelength observations of GRB, to study their astrophysics and to use them as cosmological probes. Furthermore in 2009 ECLAIRs is expected to be the only space borne instrument capable of providing a GRB trigger in near real-time with sufficient localization accuracy for GRB follow-up observations with the powerful ground based spectroscopic telescopes available by then.*
*A "Phase A study" of the ECLAIRs project has recently been launched by the French Space Agency CNES, aiming at a detailed mission design and selection for flight in 2006. The ECLAIRs mission is based on a CNES micro-satellite of the "Myriade" family and dedicated ground-based optical telescopes. The satellite payload combines a 2 sr field-of-view coded aperture mask gamma-camera using 6400 CdTe pixels for GRB detection and localization with 10 arcmin precision in the 4 to 50 keV energy band, together with a soft X-ray camera for onboard position refinement to 1 arcmin. The ground-based optical robotic telescopes will detect the GRB prompt/early afterglow emission and localize the event to arcsec accuracy, for spectroscopic follow-up observations.*

## *I. Introduction*

Gamma-ray bursts (GRB) are detected by space borne gamma-ray telescopes as an important count-rate increase during a short period of time. These phenomena were discovered in the 1960's, however their origin remained mysterious for a long time. Only in the 1990's, thanks to the BATSE detector onboard the NASA CGRO satellite, it was possible to get a first indication that those events were of cosmological origin. BATSE [1] showed that GRB are distributed uniformly on the sky, furthermore that about 80% of all GRB last less than one minute, and that there is a distinct class of short-duration GRB which last less than a few seconds. The real breakthrough in understanding GRB occurred in 1997 in the era of the Beppo-SAX satellite with the discovery of X-ray [2], optical [3], and radio [4] GRB afterglows, allowing the detection of the host galaxies and the measurement of their redshift. Those observations showed that (at least long-duration) GRB are events related to explosions of massive stars, taking place at cosmological distances. In the most commonly accepted model, the gamma-rays of a GRB are produced by internal shocks taking place in a collimated jet of particles produced during the event and directed close to the line of sight of the observer, while the afterglow results from the interaction of the jet with the surrounding medium; however alternative models exist. The energy radiated in gamma-rays by a GRB amounts to >$10^{51}$ erg (while the energy released by a typical supernova is about $10^{49}$ erg), thus GRB are the most energetic explosions known to take place in the Universe since the Big Bang.

Future studies of GRB will allow to better constrain the physical GRB models by understanding the physical processes involved in those extreme events. Of particular interests are the mechanisms capable of producing the high gamma-ray flux, the origin of the prompt emission in X-rays and the visible band, and the formation of the afterglow. Additionally the study of GRB offers important perspectives for the progress of cosmology. Due to their high intrinsic luminosity, GRB allow to probe the Universe at very high redshifts, the properties of the host galaxy and the intergalactic medium on the line of sight being imprinted in the spectrum of the GRB afterglow. The star formation rate at high z can be studied using GRB related to massive star explosions, and it could even become possible to detect the first generation of stars, responsible for the re-ionization of the Universe after the Big Band.

After its launch foreseen in 2009, the micro-satellite mission ECLAIRs (for *"flash of lightning"* in French) will detect about 100 GRB per year, observe them simultaneously in the visible and the X/γ-ray domain, and will contribute to the rich field of research which emerged from GRB observations.

## *II. ECLAIRs mission concept*

### A. GRB detection and observation strategy

The GRB observation strategy is based on multi-wavelength observations, which follow 3 steps : (i) detection of the GRB using wide field-of-view space-borne gamma-ray telescopes; near-realtime localization of the event on the sky with a few arc-minutes accuracy; (ii) observation of the prompt/early afterglow emission with robotic telescopes which point their field-of-view quickly onto the space-given error-box within tens of seconds after the event; localization with arc-seconds accuracy; (iii) spectroscopic observations of the GRB afterglow with large ground based telescopes and large space-borne X-ray telescopes.

This observation scheme is at work presently with HETE-2 [5], INTEGRAL [6] and Swift [7] (among others) in orbit, followed by event localization with small robotic telescope like REM [8] and TAROT [9] and spectroscopic studies after hours e.g. at the VLT (European Southern Observatory, Chile) and others. Faster spectroscopy of GRB afterglows, within minutes after the event, in the visible and near-infrared band is the goal of X-shooter [10], a second generation instrument to be installed in 2008 at one of the 8.2 m telescopes of the VLT. However, if not in 2008, then early in the next decade, all the space-borne GRB triggers available today will cease to function.

At this point ECLAIRs will be the future mission capable of providing frequent and fast GRB triggers with a good localization accuracy for ground-based GRB follow-up observations. Furthermore ECLAIRs will be operational simultaneously with GLAST [11]. For GRB observed with both satellites, a very large spectral coverage will therefore be available: from 1 keV by ECLAIRs up to 30 MeV by the GLAST Burst Monitor (GBM) and a few hundreds of GeV by the GLAST Large Area Telescope (LAT).

## B. The ECLAIRs system

The ECLAIRs system is composed of a micro-satellite and a dedicated ground segment. The ECLAIRs micro-satellite (Fig. 1) is of the successful Myriade family, developed by the French space agency CNES, from which up to now 6 are in orbit, DEMETER launched in June 2004 being the first one. In 2009, ECLAIRs could be launched as a passenger of the French-Indian Megha-Tropiques mission by an Indian PLSV launcher and injected into a circular low-Earth orbit below the Earth radiation belts (altitude 860 km, 20° inclination). The altitude will then be lowered to 670 km in order to reduce the influence on the detectors of the charged particles trapped in the South Atlantic Anomaly of the Earth magnetic field. The total mass of the ECLAIRs satellite is about 150 kg, including a payload mass of 65 kg.

The payload module comprises a gamma camera (CXG, *Caméra X et Gamma*), detecting photons of energy between 4 keV and >300 keV. With its imaging capability between 4 keV and 50 keV, this camera is the prime instrument used for GRB detection and localization on the sky (with 10 arcmin accuracy). As a second instrument on-board, the *Soft X-ray Camera* (SXC) provides photon detection between 1 keV and 10 keV and a refined localization capability (with 1 arcmin accuracy) in a Vernier mode, i.e. using the error box provided by the CXG in order to further refine the position determination. This refined position will allow *prompt* ground based searches of high redshifted GRB by large telescopes equipped with infra-red imagers whose field-of-view is <3 arcmin.

The ground segment of ECLAIRs is composed of the communication network and robotic telescopes in the visible and near infra-red band (UDV, *Unité de Détection dans le Visible*) dedicated to the mission. In low-Earth orbit, the satellite has no permanent high bandwidth telemetry stream with the ground, due to the lack of ground stations coverage. In order to be able to transmit GRB alerts as quickly as possible to the community of observers, a low bandwidth VHF stations network will be used. Such a network is already in place along the equator with 14 radio receiver stations connected to the internet, and is used for reception of GRB alerts from HETE-2. It is foreseen to upgrade this network (to 600 bit/s data rate) and to extend it to a total of 29 receiver stations. This broader coverage is necessary due to the 20° inclination orbit of ECLAIRs compared to HETE-2 (4°). Alerts sent through the VHF network are forwarded to the *GRB Coordinates Network* (GCN) and to the UDV. These dedicated robotic telescopes will automatically position their 30 arcmin wide field-of-view onto the position of the GRB alert and could determine the position of the possible prompt emission/early afterglow with 1 arc-second accuracy. Since an UDV telescope can only observe those candidates occurring above the horizon and during night at the UDV site, the UDV sites are located in the tropical zones and the observations program of the satellite is optimized in order to maximize the number of follow-ups.

## C. The ECLAIRs scientific specifications

The main scientific specifications of the ECLAIRs mission are summarized as follows: (i) detect about 200 GRB over the duration of the entire mission (minimum mission life 2 years), independently of their duration, detect in particular X-ray rich GRB and X-ray flashes; (ii) compute in near-realtime (10 seconds) the position of the GRB on the sky with an accuracy of 10 arcmin, and for 50% of them with a refined 1 arcmin accuracy; transmit this information to the ground based telescopes within 1 minute. (iii) estimate in 20% of the cases within 5 minutes the position of the GRB optical afterglow with a 1 arcsec accuracy. (iv) coordinate the pointing strategy with ground-based large spectroscopic telescopes in order to allow follow-up observations in 75% of the cases.

ECLAIRs is in particular optimized for redshifted cosmological GRB. Therefore the ECLAIRs X/gamma-ray detectors cover the energy band from 1 keV to ~300 keV (with GRB detection and localization to 10 arcmin in the 4 keV to 50 keV energy band, and refined 1 arcmin localization in the 1 to 10 keV band), and the UDV operates in the near infra-red band.

## *III. ECLAIRs mission components*

### A. The gamma camera (CXG)

The CXG is a wide field-of-view coded-mask aperture telescope, based on the experience gained on the INTEGRAL imager IBIS. The CXG is used both for GRB localization with 10 arcmin accuracy and photon-by-photon acquisition. Its detection plane (DPIX), developed by CESR, Toulouse, France, is made of 6400 CdTe crystals (crystal size $4\times4$ mm$^2$, thickness 1 mm, surrounded by a guard ring, and using Schottky contacts), capable to detect X-ray and $\gamma$-ray photons with high efficiency (~ 100% below 50 keV) and good spectral resolution. These crystals are coupled to low-noise readout ASICs called IDeF-X [12] (32 crystals connected per ASIC), developed at CEA Saclay, France, in the framework of a CEA-CNES R&D program. Tests with preliminary versions of IDeF-X coupled to CdTe crystals in the lab and operated at room temperature show that the low energy threshold obtained is well below 4 keV, meeting the ECLAIRs scientific requirements. The energy resolution (920 eV at 60 keV) is excellent and competes with cryogenic Ge detectors in this energy range. The CXG detection plane is segmented into 8 modules of 25 groups of 32 crystals, has a mass of 18 kg and a total power consumption of 40 W (among which 3 mW per ASIC input channel connected to a CdTe pixel). It is passively cooled to -20°C by 2 radiators. Its readout electronics uses 16 ADC (12 bit, 10 µs dead-time) and produces photon-by-photon data coded on 4 Bytes (energy, time and pixel address) sent to the data acquisition system.

The design of the CXG has a (partially coded) total field-of-view of 2 sr (88°×88°). The detection plane (80×80 pixels, 4.5 mm side each) has a sensitive area of 1024 cm$^2$, and is placed 46 cm below a coded mask (54×54 cm$^2$). The mask is made of a tungsten sheet with 30% open pixels (square holes), following a pseudo-random pattern chosen to have acceptable mechanical properties and a good self-correlation function. The CXG is surrounded by a passive shield (opaque up to 50 keV) made of successive Ta, Sn, Cu and Al foils (listed inwards), one foil shielding the X-ray fluorescence of the previous one. With such a camera, the sky is segmented at least into 199×199 pixels with an angular size of 34 arcmin (central pixel) to 18 arcmin (edge); the point source localization capability reaches 10 arcmin after fitting the mask point-spread function in case of a favorable signal-to-noise ratio. A Monte-Carlo simulation by O. Goded [12] shows that the sensitivity in the 4-50 keV band is 170 mCrab for a 10 s exposure (and for a signal-to-noise ratio of 5.5) and that the ECLAIRs GRB detection sensitivity exceeds that of Swift for a GRB with a peak energy below 40 keV (as expected for highly redshifted cosmological GRB), because of its low-energy threshold.

### B. The soft X-ray camera (SXC)

The SXC has been added to the payload since our recent paper [13] in order to improve the ECLAIRs localization accuracy in X-rays. The SXC is developed by MIT, Cambridge (MA), USA, and is based on the heritage of MIT X-ray cameras on ASCA, Chandra, HETE-2 and Astro-E. The ELAIRs SXC is made of 4 unidirectional modules, 2 modules sensitive in the Y and 2 in the Z directions. Each module observes the same field-of-view as the CXG and is equipped with a coded mask placed 10 cm above a CCD. The CCD (MIT/Lincoln Lab CCDID-41, 1K×1K pixels, 15 µm pixel size, 96 cm² sensitive area, passively cooled to -50°C) detect X-rays in the 1 keV to 10 keV energy range. The CCD are read out individually for each photon at a 10 Hz rate. The energy resolution at 5.9 keV is 132 eV; after 2 years in orbit it is expected to degrade to 210 eV due to radiation damages in the CDD. These can be purged via a charge injection technique, which restores the energy resolution to 144 eV. Each CCD is protected against micrometeorites by a 25 µm thick beryllium sheet, which reduces the puncture rate to 0.025 impacts per year per CCD. The mask of each module is made of a stress-free gold sheet, 25 mm thick, which is electroformed in order to obtain mask features accurate to 2 µm over the mask dimension of 10 cm. The mask is mounted in a stainless steel frame at high temperature such that a stabilizing tension is applied to the mask after cooling. The SXC power consumption is 4 W, its mass is 4.5 kg, and its output data rate is 40 kbit/s max.

The ECLAIRs SXC has a sensitivity of 190 mCrab for 10 s accumulation (for a signal-to-noise ratio of 4) and a source localization radius of 24 arcsec (at 90% C.L.) for a 5 sigma detection. Main issues for a good source localization are temperature stability, in order to avoid structure deformations, and a good stability of the star-tracker based attitude control.

## C. The on-board GRB trigger and data acquisition system (UTS)

The ECLAIRs on-board real-time system UTS (for *Unité de Traitement Scientifique*), developed by CEA, Saclay, France, acquires the scientific data from the CXG and SXC and sends it to the on-board mass-memory. In parallel, using the CXG data, it performs the detection of a GRB event by searching for a count rate increase or the appearance of a new gamma-ray source in a sky image. It computes the source position on the sky and sends the GRB alert message to the VHF transmitter. It also requests from the SXC the refined GRB position and transmits this one in the VHF alert message as well.

The data from the CXG are read-out photon-by-photon in parallel from each of the 16 ADC by the UTS at a maximal photon rate of 1.6 MHz (51 Mbit/s) and a mean (background) rate below 10 kHz (320 kbit/s). All received photons are accumulated into packets of fixed size, dated by Universal Time (UT), delivered by the Myriade platform, and dumped into the on-board mass-memory (16 Gbit in size). The SXC data are similarly acquired by the UTS and sent to the mass-memory (40 kbit/s max). Whenever the satellite is in sight of a high-bandwidth ground-receiver (at least twice per day), the mass-memory data are sent to ground via an X-band transmitter (at 16 Mbit/s). The payload configuration is operated through a bi-directional satellite command/control stream based on S-band receiver/transmitters (contact available a few times per day).

For the GRB detection by the count rate increase trigger, each photon received in the UTS is fed into light-curves (histograms rolling with time). In order to account for the spread in GRB duration, time histograms are built with at least 6 different binnings (ranging from 10 ms to 10.24 s), in 3 different energy bands (e.g. 4-10-20-50 keV) and 9 different regions of the detection plane (entire plane, 4 halves sliced in 2 directions, 4 quadrants). Those histograms are continuously monitored for a count-rate increase. When an increase (e.g. 5 sigma over the background estimated from the early part of the histogram) is found, the imaging software produces a sky image by de-correlating the mask pattern with the corresponding detector hitmap. A significant point source is searched in the sky image after subtraction of a scaled background sky-image built from seconds earlier. If the source location does not to coincide with a known source (using an on-board catalog) the information is forwarded to the alert generator.

For GRB detection by imaging (useful for slowly increasing light curves or long lasting cosmological GRB), sky images are constructed on time-scales ranging from 10 s to 10 min, which are continuously searched for new sources (not present in the catalog) and forwarded to the alert generator.

In case of an alert, the VHF message contains the position of the GRB in the CXG and the SXC fields-of-view with their error boxes, the time of the event in UT, the satellite pointing direction and a descriptor of the event type. In case of no alert, the VHF message is filled with housekeeping information acquired by the UTS (instrument status, count rates, spacecraft pointing direction, and information on the previous GRB alert repeated to prevent loosing it in case a VHF ground station couldn't catch this previous alert).

In case of an alert, the mass-memory zone corresponding to 5 minutes prior and 10 minutes after the event is marked as secured: it cannot be overwritten before its down-link via X-band is successfully completed. A confirmation to free this memory zone is uploaded to the payload via the S-band.

## D. The visible-band telescope (UDV)

The scientific requirements and the concept of the UDV (for *Unité de Détection dans le Visible*) has been modified since our recent paper [13]. Indeed it is extremely difficult to build a space-borne or ground-based wide field-of-view camera which reaches, in 10 s observation time, a sensitivity better than magnitude 14, as required to detect the expected prompt emission of GRB. However, a ground-based fast robotic UDV telescope system is feasible. With a smaller field-of-view and a larger sensitivity it can (i) still observe the prompt emission of the longest GRB if fast enough, (ii) localize the GRB to better than 1 arcsec within 5 minutes and provide this information to large spectroscopic telescopes, and (iii) acquire the GRB optical light curve simultaneously in different bands, from visible to infra-red, providing a magnitude estimate, and a photometric redshift determination.

The requirements on the UDV are therefore the following: (i) point each telescope automatically to the position of an ECLAIRs GRB alert in less than 10 s for 20% of all alerts, (ii) reach a magnitude 17 sensitivity in 10 s in the J band within a field-of-view of 0.5°×0.5°, compatible with the CXG error box, (iii) perform simultaneous observations in the visible (R,J bands) and the near infra-red (J, H and Ks bands) for the redshifted cosmological GRB, (iv) setup the telescopes near large spectroscopic observatories (e.g. X-Shooter) in order to optimize their follow-up possibilities. Such UDV will use standard commercial components. Dedicated to ECLAIRs, they will be tolerant to an increased false-trigger rate, and provide a privileged data access for the ECLAIRs community.

## *IV. Concluding remarks*

The ECLAIRs mission follows a CNES micro-satellite development plan (reduced cost, minimal number of intermediate models before the flight-model, high-grade commercial components) and aims at a 2 years lifetime. The mission is currently in phase-A since Mai 2005 and aims at a detailed mission definition in early 2006. The decision for construction of ECLAIRs will take place in late 2006. The development of the subsystems is planned between 2006-2007, the mission integration and ground-based tests are foreseen in 2008. The launch is scheduled in 2009. From this moment on, the ECLAIRs satellite will provide GRB triggers and follow-up observations for the GRB community.

*Acknowledgements*

The authors would like to thank very much M.-A. Clair, Ph. Lier, M. Bach, V. Cipolla, Th. Chapuis and colleagues from the CNES Micro-Satellite Division in Toulouse, France, for their intense contribution and involvement in the ECLAIRs project.

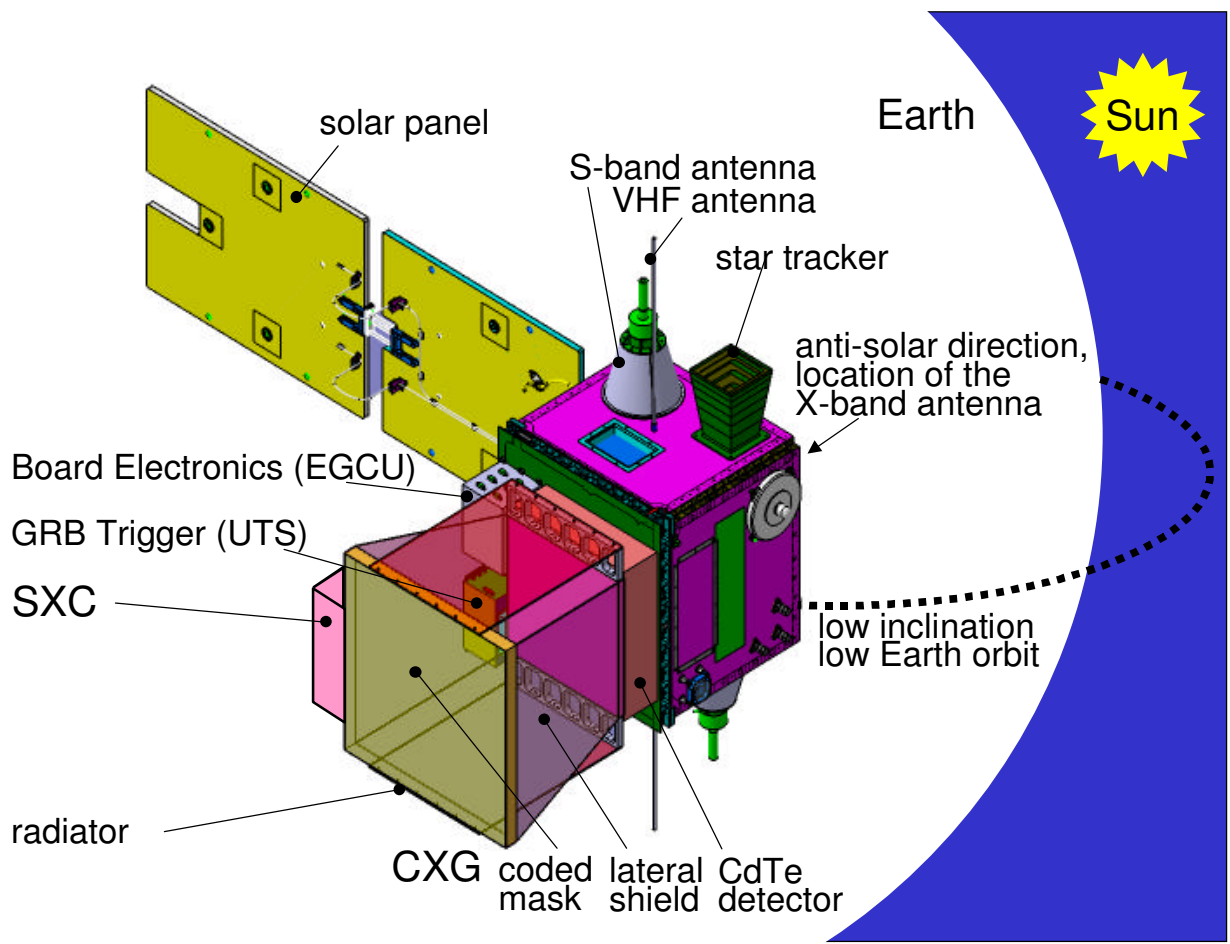

Fig. 1: This schematic view of the ECLAIRs micro-satellite shows the cubic shaped Myriade micro-satellite platform (including the solar panel, S-band, X-band and VHF transmitters and the star tracker; dimensions of the cube ~ 60 cm), and the ECLAIRs payload, including the X/gamma camera (CXG) and the Soft X-ray camera (SXC), the scientific trigger electronics (UTS) and the payload board electronics (EGCU).